\documentclass[12pt]{iopart}
\usepackage{graphicx}
\usepackage{array}
\usepackage{latexsym}
\usepackage{mathrsfs}
\usepackage{color}
\usepackage{hyperref}
\usepackage{iopams}

\newcommand{\bra}[1]{\langle #1|}
\newcommand{\ket}[1]{|#1\rangle}

\newcommand{\abs}[1]{\left|#1\right|}
\newcommand{\ave}[1]{\left<#1\right>}

\newcommand{\openone}{\leavevmode\hbox{\small1\normalsize\kern-.33em1}}

\begin{document}

\title[Entanglement indicators for quantum optical fields]{Entanglement indicators for quantum optical fields: three-mode multiport beamsplitters EPR interference experiments}

\author{Junghee Ryu$^{1,2}$, Marcin Marciniak$^2$, Marcin Wie\'{s}niak$^{2,3}$ and Marek \.{Z}ukowski$^2$}
\address{$^1$ Centre for Quantum Technologies, National University of Singapore, 3 Science Drive 2, 117543 Singapore, Singapore}
\address{$^2$ Institute of Theoretical Physics and Astrophysics, Faculty of Mathematics, Physics and Informatics, University of Gda\'{n}sk, 80-308 Gda\'{n}sk, Poland}
\address{$^3$ Institute of Informatics, Faculty of Mathematics, Physics and Informatics, University of Gda\'{n}sk, 80-308 Gda\'{n}sk, Poland}
\ead{rjhui82@gmail.com}

\begin{abstract}
We generalize a new approach to entanglement conditions for light of undefined photons numbers given in [Phys. Rev. A {\bf 95}, 042113 (2017)] for polarization correlations to a broader family of interferometric phenomena.
Integrated optics allows one to perform  experiments based upon multiport beamsplitters. To observe entanglement effects one can use multi-mode parametric down-conversion emissions. When the structure of the Hamiltonian governing the emissions has (infinitely) many equivalent Schmidt decompositions into modes (beams), one can have perfect EPR-like correlations of numbers of photons emitted into ``conjugate modes" which can be monitored at spatially  separated detection stations. We provide entanglement conditions for experiments involving three modes on each side, and three-input-three-output multiport beamsplitters, and show their violations by bright squeezed vacuum states. We show that a condition expressed in terms of averages of observed rates is a much better entanglement indicator than a related one for the usual intensity variables. Thus the rates seem to emerge as a powerful concept in quantum optics, especially for fields of undefined intensities.
\end{abstract}


\maketitle

\section{Introduction}

New methods of correlation analysis in  multi-photon quantum optics were introduced in \cite{ZUKOWSKI15b} and \cite{ZUKOWSKI15a}. They allow to construct much better entanglement witnesses, and new more effective Bell inequalities. The inequalities are devoid of theoretical loopholes (i.e. additional assumptions, apart from local-realism/causality). 
What is very important  generalized methods were devised, which allow to find other entanglement conditions, beyond the ones explicitly derived in the papers.  The ``headline'' result of this line of research is the new approach to polarization correlations of quantum light for states of undefined intensities.  This leads to a revision of the concept of quantum Stokes observables.

In \cite{ZUKOWSKI15a}, one can find a derivation which shows that the standard “textbook” Bell inequalities \cite{REID, WALLS} for quantum optical fields, which address the case of 
states of undefined intensities (or for simplicity, photon numbers), can be replaced by new ones, which do not rest  on certain additional assumptions, which were necessary to derive the textbook ones. To this end, one has to use redefined observables, which we shall call rates.  In the case of polarization measurements (say discriminating between H and V polarized photons) such a rate registered at by a  detector in, say, channel H is  the ratio of  the number of photons registered by it, to the number of photons counted by both detectors (for a given run, {\em not} averages). The eigenvalues of such rate observables are rational numbers between one and zero. Because of that,  in a smooth way one can re-write any known Bell inequalities for pairs of particles, to get Bell inequalities for optical fields  in terms of the rates \cite{ZUKOWSKI15a}. It is worth noting that the new inequalities, which do not require additional assumptions, can detect entanglement in situations in which the “standard” ones \cite{REID, WALLS} fail.

The above results lead  to a reconsideration of the usage of the Stokes parameters for quantum optical fields \cite{ZUKOWSKI15b}. Despite the fact that the parameters  were introduced in 1852(!), they are used in quantum optics without any modification. They are the differences of the average intensities (or photon numbers) of light exiting polarization analyzers, measured in three complementary arrangements (horizontal-vertical, diagonal-antidiagonal and right-left-handed circular polarization analysis), and the total average intensity. If the photon numbers are undefined, the instances when their high values are registered contribute more to the parameters. Redefined quantum Stokes parameters, introduced in \cite{ZUKOWSKI15b},  describe the averaged measured polarization with influence of intensity fluctuations removed. Note that the quantum state describes a statistical ensemble of equivalently prepared systems. The average value of an observable is taken over such a statistical ensemble. In an experiment, this implies many repetitions (runs) with the average of the results of the runs giving the experimental value of the observable. Each Stokes observable is redefined  as the ratio of the difference of photon numbers at the two exits of a polarizing beam splitter to their sum. It is this ratio that is to be averaged, both in the experiment and theory, and not the numerator and the denominator separately, as in the conventional approach. As a result, in each run the registered intensity of light does not influence the weight of a given run in calculating the average polarization. This approach is most useful in the case of observation of polarization correlations at two or more separate detection stations. In order to measure the correlations  of  new Stokes parameters, one does not require any new registration techniques, when compared with measurements of old Stokes observables. All that is needed is a different analysis of the data.

The new  Stokes parameters allow to re-formulate entanglement conditions, so that they,  in the case of the (four-mode, bright) squeezed vacuum (i.e. the output of a strongly pumped down conversion source), allow one to detect entanglement via polarization measurements for stronger pumping and hence for larger mean photon numbers, as well as for higher losses. The approach also allows one to re-write (or map)  any  entanglement condition for two qubits in such a way that we get a condition for polarization of quantum optical fields, which employs the modified Stokes parameters. Thus we can now construct plethora of new entanglement conditions for correlations of quantum light. 
The new ideas of replacing intensities via rates in correlation functions, lead to stronger visibility of some non-classical phenomena, in the case of quantum states of undefined photon numbers, see \cite{HOM2017} in which the working example is the Hong-Ou-Mandel dip \cite{HOM87}, under strong pumping condition.

Here we extend this approach beyond polarization effects. We hope that the results will contribute to the experimental search of non-classical phenomena with integrated optics methods, which allow observations stable multichannel interference effects. With the recent progress in photon number resolved detection, our new entanglement conditions may play an important role in the field. We shall study entanglement conditions for bright multi-mode quantum optical fields of undefined intensities (essentially, photon numbers). Extensions of the approach to Bell inequalities for multiport experiments will be presented elsewhere.

The gedanken but already feasible experiments which we study  involve pairs of  spatially separated multi-port beam-splitter interferometers, of the kind studied in \cite{MULTIPORT97}. Multi-port techniques were tested by Walker \cite{WALKER86, WALKER87}. 
Ideas concerning their use to observe effects related with higher dimensional entanglement one can find in \cite{ZEILINGER93a, ZEILINGER93b}.
The fact that the multi-port interferometers can perform any unitary transformations of finite dimensional single photon states was shown by Reck \etal \cite{RECK}. In 2000, it was shown that two-photon higher dimensional entanglement leads to stronger violation of local realism than two-qubit one (numerical results of \cite{KASZLIKOWSKI}, confirmed analytically in \cite{KASZLIKOWSKI-2} and \cite{CGLMP}). Also, two-particle higher dimensional entanglement has specific traits which are not present in qubit systems \cite{HORODECKI}.

The work of Reck \etal provides an operational blueprint for any finite dimensional unitary $d\times d$ transformations of single particle states. If the description is limited of just states which allow superpositions of a (single) particle to be in a particular beam, and all other degrees of freedom are treats as irrelevant, then  the Reck \etal transformations are produced with the use suitably interconnected  beam splitters and phase shifters, forming $d$-input-$d$-output multiports. Such multi-port devices give optical beams (modes) coupling via a unitary relation between the input and output modes. If the state of a single particle (photon) of being  in  input beam $i$, where $i=1,...,d$ is denoted by ${\phi_i}^{\rm in}$, and the states of being in output beams $k=1,...,d$ are denoted by $\phi_k^{\rm out}$,  then the unitary transformation describing the action of  multi-port, in the form of transformation of the basis states is given by 
$\phi_k^{\rm out}=\sum_{i=1}^d U_{ki}{\phi_i}^{\rm in}$, where $U$ is the related unitary matrix. This implies for the second quantized description that for the transformation of  photon creation operators related to the modes   reads $a_k^{\dagger \rm out}=\sum_i U_{ki}{a_i^\dagger}$.
 Note that the basis property of the considered single photon states implies the following commutation relations $[a_i, a_j^\dagger]=\delta_{ij}$, and  $[a_i, a_j]=0$. Identical relations also hold for the `out' operators. Such devices, as they are generalizations of Mach-Zehnder  ($2\times 2$) interferometers in principle allow to experimentally/operationally test basic single and two (three, \dots, etc.) photon interference and quantum information processes, see e.g., \cite{PAN}. Here we want to study the extensions of such experiments to the second quantized optical fields, having in mind especially states of light with undefined photon numbers. We investigate non-separability criteria. As our working example  of an entangled quantum optical state, we shall take the six-mode bright squeezed vacuum. Such states allow perfect EPR correlations, by which we understand perfect correlations for at least two complementary measurement arrangements. 
 Using this property one can derive entanglement conditions, as separable states cannot have EPR correlations.
 Entanglement conditions for quantum optical files derivable from EPR correlations were given for the case of four-mode squeezed (polarization entangled) vacuum in \cite{BOUW}, see also \cite{MASZAa, MASZAb}. They are in the form of conditions for correlations of standard polarization measurements, that is for Stokes parameters. One can easily generalize the EPR correlations to higher number of modes, see further.

With the ongoing improvements in parametric down-conversion techniques, the birth of integrated optics, and laser imprinting methods to build such devices, the multi-port interferometry experiments, such as the ones suggested in \cite{MULTIPORT97}, are becoming feasible. As a matter of fact,  important tests of exactly such configurations were recently done, see Schaeff \etal \cite{ZEIL-2015}. The schemes discussed here involve   parametric down-conversion for higher pump powers, in the case of which we do not have only spontaneous emissions of pairs of correlated photons, but superpositions of multi-pair emissions.
Therefore new phenomena need to be studied. At least one should check to what extent the features of two-photon correlations are still present in the case of stronger fields.

We take as our example a six-mode squeezed vacuum state which can be produced with the use of a parametric down conversion crystal.
However, our intention is not a proposal of feasible experiment, but introduction of new entanglement conditions
for multimode optical states of undefined photon numbers. The operational situation which we study serves only as an example here.
With the current rapid development of integrated optics, which now includes not only integrated multimode interferometers
\cite{Weihs96, Meany12, Metcalf13, Spagnolo13, CAROLAN15, Peruzzo11}, but also sources \cite{Tanzilli01, Takesue05, Martin10, Jin14, Matsuda12, Herrmann13, Silverstone14}, the schematic configuration which we present may become feasible.

We shall consider such sets of settings of the local multiport interferometers,  which in the case of single photons allow transformations to full sets of mutually unbiased state bases. We shall limit here our considerations to the first non-trivial step, that is to $d=3$. Higher `dimensional' experiments of a similar kind  have been studied in~\cite{RYU18}.
As a by-product of our considerations we shall also get complementarity relations for $3\times3$ multi-port interferometry of general quantum optical fields. We shall see that many of the traits of complementarity relations for single particle multi-port measurements still hold for fields of undefined photon numbers.  If one replaces intensities at the outputs of such devices by rates (ratios between the observed intensity at the given output divided by the total intensity), such relations are quite elegant.  

The broader implication of our results is that they question the usual paradigm that the quantum coherence properties of optical fields can be best revealed by intensity correlation functions, see {\em any} textbook of Quantum Optics. The results presented here show that, in the case in which one can use correlations between rates, instead of the usual intensity correlations, one often gains in 
the visibility of non-classical phenomena.  This finding will be additionally supported in forthcoming publications.

\section{ EPR correlations: sources, states and measurements}

 The entanglement indicators of Ref. \cite{ZUKOWSKI15b} are for polarization correlations. They involve  measurements  of three mutually unbiased, complementary polarizations: e.g.,  horizontal/vertical, diagonal/anti-diagonal and circular right/left handed. Here, we shall derive generalizations of such entanglement conditions for  the multi-mode cases. 
 
 The construction of entanglement indicators of \cite{BOUW} and \cite{ZUKOWSKI15b} takes as its starting point the fact that for the four-mode BSV one can observe perfect EPR correlations (in many pairs of polarization measurements bases), and that separable states do not have this property. They can be only classically correlated. We shall extend this idea to the multimode case. 
 
Here, as our starting point we take multi-mode emissions in the down-conversion process \cite{MULTIPORT97, PAN}. The emissions from the parametric down-conversion source are directionally correlated due to the phase matching conditions. In the type-I parametric down-conversion, the pairs of photons of the same frequency are emitted into a cone, in such a way that one can register coincidences into pairs of directions along the cone which lay in the same plane as the pump field, for details see \cite{PAN}. One can select in principle several  pairs of such directions, and collect their radiations.

The interaction Hamiltonian which describes such an arrangement has the following form:
\begin{eqnarray}
H= \rmi \gamma \sum_{i=0}^{d-1} a_{i}^{\dagger} b_{i}^{\dagger} + h.c.,
\label{EQ:H_INT}
\end{eqnarray}
where $a_{i}^{\dagger}$ and $b_{i}^{\dagger}$ are the creation operators of $i$th signal-idler mode pair, and $\gamma$ is a coupling constant proportional to the pumping power. The modes $a_i$ are directed (via optical fibers, etc.) to a detection station `Alice', while modes $b_i$ to `Bob'. Notice that the Hamiltonian
can be put in the following form: $ \rmi \gamma \sum_{i,j=0}^{d-1} \delta_{ij} a_{i}^{\dagger} b_{j}^{\dagger} + h.c.$ If one takes a $d\times d$
unitary matrix $U$, one has $\delta_{ij}=\sum_kU^*_{ki}U_{kj}$. Further if one defines $a_k^{\dagger \rm out}=\sum_j U_{kj}{a_j^\dagger}$, and 
$b_k^{\dagger \rm out}=\sum_i U^*_{ki}{b_i^\dagger}$, then one can write the Hamiltonian down in an equivalent form:
\begin{eqnarray}
H= \rmi \gamma \sum_{k=0}^{d-1}  a_{k}^{\dagger \rm out} b_{k}^{\dagger \rm out} + h.c.
\end{eqnarray}
 This symmetry of $H$ implies an invariance of the perfect EPR correlations.
Such a transformation can be done using a specific pairs of `conjugate' multi-port interferometers, one for Alice one for Bob. In other words, as the squeezed vacuum state resulting from the application of Hamiltonian $H$ on initial vacuum reveals perfect correlations, such correlations also occur after the pair of local  mode transformations. 

Notice that one can consider the $U$ transformations matrices of modes which are associated with unitary transformations leading to unbiased bases for a $d$-dimensional Hilbert space. 

\subsection{Example}

Consider $d=3$.
For three pairs of Schmidt modes, the emitted photon pairs are prepared in the following entangled state:
\begin{eqnarray}
\ket{{\rm BSV}}=\frac{1}{\cosh^3 \Gamma} \sum_{n=0}^{\infty} \sqrt{\frac{(n+1)(n+2)}{2}} \tanh^n\Gamma \ket{\psi^{n}},
\label{EQ:BSV}
\end{eqnarray}
where
\begin{eqnarray}
\ket{\psi^n}=\sqrt{\frac{2}{(n+1)(n+2)}}\sum_{p_0+p_1+p_2=n} \ket{p_0}_{a_0} \ket{p_1}_{a_1} \ket{p_2}_{a_2} \ket{p_0}_{b_0} \ket{p_1}_{b_1} \ket{p_2}_{b_2}.
\label{EQ:BSV_2}
\end{eqnarray}
Here, the sum is taken over all combinations of nonnegative integers $p_i$. The parameter $\Gamma$ describes `gain' and is dependent on 
$\gamma$ and the interaction time (basically equal the length of the non-linear crystal divided by the speed of light).

\begin{figure}[h]
\center
\includegraphics[width=12cm]{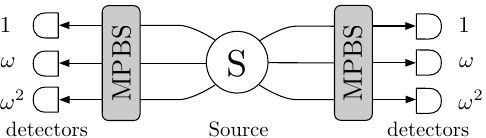}
\caption{Schematic diagram of the measurement devices for the three pairs of Schmidt modes. They consist of three-input-three-output multiport beamsplitters (denoted by MPBS) and detectors. By the interaction Hamiltonian (\ref{EQ:H_INT}), the six-mode bright squeezed vacuum state (\ref{EQ:BSV}) is generated. This entangled state leads to  perfect EPR correlations between the local conjugate modes. }
\label{FIG:state_d3}
\end{figure}

Our measurement devices consist of an unbiased symmetric multi-port beam-splitter \cite{MULTIPORT97} and detectors behind the beam splitters. The unbiased multi-port beam-splitter is defined as an $d$-input and $d$-output interferometric device, of the property that  light entering  via only a single port is split to all output ports, $1/d$ of the intensity into each exit. Following the one photon case  \cite{MULTIPORT97} one can relate with each exit $d$th complex roots of unity, values $\omega=\exp(2 \pi \rmi /d)$. The three-output case is shown in figure \ref{FIG:state_d3}.

\section{Entanglement criteria for three-port beamsplitters EPR interference}
\label{SEC:W_ENT}

We shall study here an approach which uses the specific value assignment as in figure \ref{FIG:state_d3} (see \cite{MULTIPORT97}). Consider four unitary transformations which lead to the unbiased (complementary) bases in a $3$-dimensional Hilbert space. Note that generally when the dimension $d$ of a Hilbert space is an integer power of a prime number, the number of mutually unbiased bases is known to be $d+1$ \cite{WOOTTERSa, WOOTTERSb}. We put $U(3)=\openone$, while the other three, indexed with $m=0,1,2$, have matrix elements which lead to the following transformations of the bases \cite{WOOTTERSa, WOOTTERSb}: 
\begin{eqnarray}
U(m)_{js} = \frac{1}{\sqrt{3}} \omega^{js +ms^2},
\end{eqnarray}
where now $\omega=\exp(2 \pi \rmi /3)$.
With such transformations one can relate a multi-port beam-splitter (interferometer) which couples the input beams the creation operators, $a_i^\dagger$, with the output ones, $a_{j}^{\dagger} (m)$ in the following way: 
\begin{eqnarray}
a_{j}^{\dagger} (m) = \frac{1}{\sqrt{3 }} \sum_{s=0}^{2} \omega^{js +ms^2} a_{s}^{\dagger},
\end{eqnarray}
and we define $a_{j}^{\dagger} (m=3) =  a_{j}^{\dagger}$.

 As our (local) observables we shall define, for each of the four complementary (local) multiports  $m$,
\begin{eqnarray}
\hat{R}_m=\hat{r}_{0}(m) + \omega \hat{r}_{1}(m) + \omega^{2} \hat{r}_{2}(m),
\label{EQ:MO}
\end{eqnarray}
where $ \hat{r}_{j}(m) \equiv \Pi \, \frac{\hat{n}_{j}(m)}{\hat{N}} \, \Pi $  is the {\em rate operator} for exit mode  $j$ of multiport $m$. In the formula for the rate $\hat{n}_{j}(m)=a_{j}^{\dagger}(m) a_{j}(m)$ is the photon number operator for the mode. The symbol   $\hat{N}=\sum_{j} \hat{n}_{j}(m)$ stands for the operator of the total number of photons (it is invariant with respect to unitary mode transformations). Finally, we have the projection operator  $\Pi = \openone -\ket{\Omega}\bra{\Omega}$, where  $\ket{\Omega}$ is the vacuum state of the three modes. Thanks to the use of the above,  the operator $\hat{R}_m$ acts only in the non-vacuum part of the Fock space of photons. This trick makes the operator ${1}/{\hat{N}}$ properly defined.

As the squeezed vacuum state has EPR correlations in measured numbers of photons, therefore one has
\begin{eqnarray}
\sum_{m=0}^3 \ave{ \abs{ \hat{R}_{m}^{A} -  \hat{R}_{m}^{B}}^2}_{{\rm BSV}} = 0,
\label{EQ:zero_cond}
\end{eqnarray}
where indices $A,B$ mark operators  for Alice and Bob, respectively. This notation will be used whenever we want to distinguish the two local situations.  

We will show that for separable states  the condition (\ref{EQ:zero_cond}) does not hold. Instead of it, one has
\begin{eqnarray} \label{EQ:ENT_COND}
\sum_{m=0}^{3} \ave{ \abs{ \hat{R}_{m}^{A} - \hat{R}_{m}^{B} }^2 }_{\rm sep}
 \geq& 3 \left( \ave{\Pi^{A} \frac{1}{\hat{N}^{A}} \Pi^{A}}_{\rm sep} + \ave{\Pi^{B} \frac{1}{\hat{N}^{B}} \Pi^{B}}_{\rm sep} \right).
\end{eqnarray}
To this end, we use the following two formulas. In \ref{APX:derivation}, we show the following operator relation
\begin{eqnarray}
\sum_{m=0}^3 \abs{\hat{R}_{m} }^2 = \Pi + \Pi \frac{3}{\hat{N}}\Pi, 
\label{EQ:STOKES_2}
\end{eqnarray}
while in \ref{APX:UBOUND_R} we also show that for any separable state, $\varrho_{\rm sep}=\sum_k p_k \varrho_{k}^{A} \otimes \varrho_{k}^{B}$, one has
\begin{eqnarray}
\sum_{m}\ave{\hat{R}_{m}^{A}\hat{R}^{\dagger B}_{m} + \hat{R}_{m}^{\dagger A}\hat{R}_{m}^{B}}_{\rm sep} \leq 2.
\label{EQ:LOWER_B}
\end{eqnarray}
The condition is  {\it necessary} for a state to be separable.

It is easy to check, just by retracing our calculations in the appendices, that an analog condition involving photon numbers rather than rates reads:
\begin{eqnarray}
\sum_{m=0}^{3} \ave{ \abs{ \hat{K}_{m}^{A} - \hat{K}_{m}^{B} }^2 }_{\rm sep} \geq 3  \ave{\hat{N}^A +\hat{N}^B}_{\rm sep},
\label{INTENSITIES}
\end{eqnarray}
where $ \hat{K}_m=\hat{n}_{0}(m) + \omega \hat{n}_{1}(m) + \omega^{2} \hat{n}_{2}(m).$ This is a (direct)  generalization of the condition of \cite{BOUW} to $d=3$, in the case of which we have summation over $m$ up to $d=2$, on the right hand side we also replace 3 by 2,
and (standard) Stokes parameters,  $ \hat{S}_m=\hat{n}_{0}(m) - \hat{n}_{1}(m)$,  replace  $\hat{K}_m$, that is we have $\omega=-1=e^{\rmi\pi}$. The index $m$ numbers three fully complementary polarization measurements.

Note, that (\ref{EQ:LOWER_B}) is by itself a separability condition. Its stronger version can be put as
\begin{eqnarray}
\sum_{m}\ave{\hat{R}_{m}^{A}\hat{R}^{\dagger B}_{m} + \hat{R}_{m}^{\dagger A}\hat{R}_{m}^{B}}_{\rm sep} \leq 2 \langle\Pi^A \rangle_{\rm sep}
 \langle\Pi^B \rangle_{\rm sep},
\label{EQ:LOWER_NEXT}
\end{eqnarray}
see \ref{APX:UBOUND_R}.

\subsection{Comparison: entanglement  conditions employing intensities vs. the ones for rates }
In \ref{APX:LOSS}, we give an analysis of noise resistance of the above  entanglement criteria. This is given for our `reference' state, the bright squeezed (six-mode) vacuum. The considered noise is the one for photon losses. We assume that all detectors in the two multi-port experiment of Figure~\ref{FIG:state_d3} are of finite efficiency, $\eta$. The critical efficiencies for the two conditions are very interesting.

\begin{figure}[h]
\centering
 \includegraphics[width=7cm]{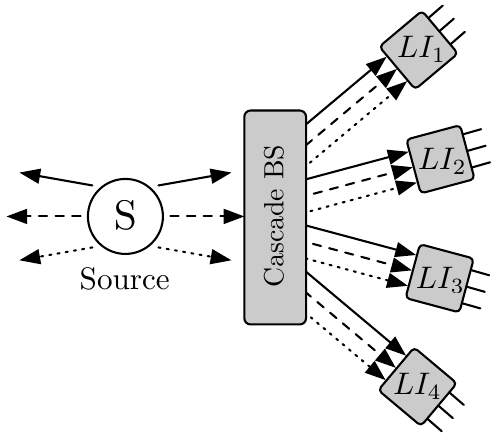}
\caption{On Alice's side each beam is split into four (branches). Each such branch is fed to a different local interferometer (LI). Each interferometer is linked with a different mutually unbiased basis (for single photons, see the main text). The same is done on the  side of Bob (not shown). 
Each original beam of Alice is split into four branches by a cascade of 50-50 beam-splitters (denoted by Cascade BS), the initial one,  and one in the reflected output of the initial one, and one in the transmitted output of the initial one, in such a way that $1/4$ of the ”intensity” always  goes to each branch.
The $LI_i$ multiport interferometers are such that each one is tuned to perform a mode transformation, related with a different mutually unbiased basis (for $d=3$).
Thus, for perfect detectors, $100\%$ efficient, counts at conjugated  interferometers of Alice,  $LI_i$ $(i=1,2,3,4)$  and of the respective contrivance  on Bob’s side (not shown), with the same  “$i$” on both sides, are such as if there was just one pair of conjugate interferometers-multiports, like in figure~\ref{FIG:state_d3}, and as if the efficiency of detection was 1/4.
Only correlations for (all) pairs conjugated interferometers of Alice and Bob are taken into account.  Thus all measurements in all pairs of conjugated bases, considered in the main text, can be done simultaneously, at the price of having a lower effective efficiency. Such an arrangement can detect entanglement only if one use the condition based on the rates. The condition with intensities is useless in this case, as its threshold efficiency is $1/4$.}
\label{FIG:d3_comp}
\end{figure}

For the criterion (\ref{INTENSITIES}) based on photon numbers, we obtain requirement of $\eta>1/4$, for {\em all} values of $\Gamma$. Notice that it is a very telling result. It means that, with even ideal detections we cannot get an experiment revealing entanglement of bright squeezed vacuum by splitting the radiation of the source on both sides into four directions (each beam) and then directing each of the branches to four pairs of conjugate complementary interferometers,  see figure~\ref{FIG:d3_comp}  (if this is unclear for the Reader, please note that in the  $d=2$  case of polarized beams, as considered in \cite{ZUKOWSKI15b} an equivalent arrangement would be to split the beam leading say to Alice into three of identical intensities, and measuring in each of the three beams, or branches, three fully complementary polarizations, and a similar action on Bob's side).  Such splitting, if one recalls the rule that a passive optical device can be permuted (if this does not lead to a different interferometric setup), acts in each branch in the same way as if the detection efficiency in the branch is $1/4$. Therefore, we see that the intensity based criterion  even for perfect efficiency inherits the usual complementarity features of single photon experiments. If one tries to make simultaneously measurements in {\em all} full complementary situations, the entanglement criterion is worthless. However, this is not so for the criterion based on the rates (\ref{EQ:ENT_COND}). For a very low gain $\Gamma$, the critical efficiency is by a whisker below $1/4$ (this is reflecting the fact that the experiment in such a regime is effectively a two-photon one, and standard Bohrian complementarity applies). But for very high  $\Gamma$ one has robustly $\eta_{\rm crit}< 1/4$, and as a matter of fact $\eta_{\rm crit}$ can be as low as $0.154$.  Thus we have not only a better resistance to losses, but additionally, in principle we can detect entanglement of very bright squeezed vacuum by beam-splitting its radiation to each side into four channels and making all the measurements at the same time.
This hints that with the use of the rates we are probing deeper into the nature  of multi-photon  light.


\section{Complementarity relations}
In Section \ref{SEC:W_ENT}, we show a separability condition based on the local operator with the specific measurement assignments (the power of $\omega$). To this end, we use the following relation (for the derivation, see \ref{APX:UBOUND_R})
\begin{equation}
\sum_{m=0}^{3} |\langle \hat{R}_{m}\rangle|^2 \leq 1.
\end{equation}
Note that this is a complementarity relation for the four possible, mutually exclusive interferometric measurements involving $d=3$ beams. The interferometers are such that they  perform unitary transformations leading to mutually unbiased bases for single photon state.
If, say $|\langle \hat{R}_{1}\rangle|=1,$ then for all $i\neq 1$ we have $|\langle \hat{R}_{i}\rangle|=0$.

\section{Summary and closing remarks}
For  $2\times 3$-mode quantum optical fields of undefined intensities, we formulate the entanglement criteria inspired by properties of EPR correlations, which are generalizations of the ones presented in \cite{BOUW} and \cite{ZUKOWSKI15b}. The first ones are based on intensities, and the second one use rates.   As an example, we consider a six-mode bright squeezed vacuum state. Such optical states have  EPR-like correlations of numbers of photons registered  in conjugated modes. With the help of multi-port beam-splitter techniques, we are able to see such correlations.   In case of inefficient detection, our approach in terms of rates is able to detect entanglement for a wider range of parameters describing the state (pumping strength) and detection efficiency.
As the critical efficiencies are quite moderate, and generation of squeezed six-mode squeezed vacuum seems feasible,   our entanglement conditions can find application in experiments. This is mainly due to the fact that integrated optics techniques allow now to produce stable multiport interferometers, this allows to put the ideas of \cite{MULTIPORT97} in practice. Such arrangements, like the one of figure \ref{FIG:state_d3} may have various quantum informational applications. These applications may go beyond single photon at single detection station paradigm, as we show that EPR correlations, which reveal non-classicality are observable also in the case of undefined photon numbers.

We have derived both conditions which are more traditional, that is based on correlation of intensities, and conditions, inspired by \cite{ZUKOWSKI15b}, which use correlations of rates. The latter ones are capable to detect entanglement where the former fail.

The other important consequence is that the presented results confirm our conjecture  that  correlation functions involving rates rather than intensities  can  become a useful tool in quantum optics, and that at least in some cases they outperform the standard ones based on intensities. We expect that one can find benefits by using the rates in various cases, e.g., quantum steering and etc.  
The results can be generalized to all $d$ for which $d+1$ mutually unbiased bases are known to exits, and other methods of detecting entanglement, see our forthcoming manuscripts.
The approach with rates is also very handy in the case of formulation of Bell inequalities, see e.g. \cite{ZUKOWSKI15a} for the case of polarization correlations..

\ack
We thank prof. M. Chekhova for discussions. This work was at its earlier stage  supported by the  EU project BRISQ2, and additionally  subsidized from funds for science for years 2012-2015 approved for international co-financed project BRISQ2 by Polish Ministry of Science and Higher Education (MNiSW). The team of authors was additionally supported by TEAM project of FNP. JR acknowledges the National Research Foundation, Prime Minister’s Office, Singapore and the Ministry of Education, Singapore under the Research Centres of Excellence programme. MZ acknowledges FNP-DFG award-grant COPERNICUS. MW acknowledges UMO-2015/19/B/ST-2/01999.



\appendix
\section{Derivation of (\ref{EQ:STOKES_2})}
\label{APX:derivation}
Let us generalize for while our considerations to prime $d$ mode case. Let $A^{\dagger}$ be a row matrix as $(a_{0}^{\dagger}, a_{1}^{\dagger}, \dots, a_{d-1}^{\dagger})$, and then its ``column Hermitian conjugate'' $A$ involves the annihilation operators. One can put $\hat{R}_m=\Pi \frac{A^{\dagger} {M}(m)A} {\hat{N}} \Pi $, where the $d\times d$ matrix
 ${M}(m)$ is given by
\begin{equation}
M(m)_{sr}=\sum_{j=0}^{d-1} \omega^{j} \bar{U}(m)_{jr}{U}(m)_{js},
\end{equation}
which is an analog of $\hat{R}_m$ for a $d$-dimensional Hilbert space (of single photon states). The matrices form the unitary  generalizations of Pauli operators for any $d$ (prime) dimensional Hilbert space.  For $d=3$ we put $M(d)_{sr}=\sum_{j=0}^{d-1} \omega^{j} \delta_{jr}\delta_{js} $, which we shall denote the matrix $M(d)$ as $Z$.  If one defines $X$ by $X_{sr} =\sum_{j=0}^{d-1} \delta_{j,s}\delta_{j+1,r}$ (where necessary, all formulas here are modulo $d$, with respect to indices), one has  $M(d)=Z$ and $M(k)=\omega^{k}XZ^{-2k}$ for $k\neq d$. The set of $XZ^{-2k}$'s  is just a permutation of the set of  $XZ^k$'s. One has $Z^0 = \openone=[M(k)]^d$. 

Using the above algebraic relations we can put: 
\begin{eqnarray}
\sum_{m=0}^d \abs{ \hat{R}_{m}}^2 = \Pi  \frac{1}{\hat{N}} A^{\dagger} \left( Z A\Pi A^{\dagger} Z^{\dagger} + \sum_{k=0}^{d-1} XZ^{k} A\Pi A^{\dagger} (XZ^{k})^{\dagger} \right) A  \frac{1}{\hat{N}} \Pi.
\end{eqnarray}
Note that the $\Pi=\openone-|\Omega\rangle\langle\Omega|$ and $\hat{N}$ commute with each other and with any operator of the form $a_j^\dagger a_i$. We make the following  transformations (on the way of which we use the following relations: $\sum_{k=0}^{d-1} \omega^{k l} = d\delta_{0l}$, $[a_i, a_j^{\dagger}]=\delta_{ij}$ and $a_i^{\dagger}a_i=\hat{n}_i$): 
\begin{eqnarray}
\fl\sum_{m=0}^d\abs{ \hat{R}_{m}}^2 &=& \Pi \frac{1}{\hat{N}^{2}}\sum_{i,j} a_i^{\dagger} \left[  \left(\omega^{i-j} a_{i} a_{j}^{\dagger} + \sum_{k=0}^{d-1} \omega^{k(i-j)} a_{i+1} a_{j+1}^{\dagger}\right)  \right] a_j \Pi, \nonumber \\
\fl &=&\Pi \frac{1}{\hat{N}^{2}}  \left( \sum_{i, j} \omega^{i-j} a_i^{\dagger} a_{i} a_{j}^{\dagger}a_j  + d \sum_{i} a_i^\dagger 
a_{i+1} a_{i+1}^{\dagger}a_i  \right) \Pi, \nonumber \\
\fl &=& \Pi  \frac{1}{\hat{N}^{2}} \left[ \sum_{i} \left( a_i^{\dagger}a_{i}  a_{i}^{\dagger}a_i + d a_i^{\dagger} a_{i+1} a_{i+1}^{\dagger}a_i \right)  + \sum_{i\neq j} \omega^{i-j}a_i^\dagger a_{i} a_{j}^{\dagger}a_j  \right] \Pi, \nonumber\\
\fl &=& \Pi \frac{1}{\hat{N}^2}\left[ \sum_{i} \left(\hat{n}_{i} \hat{n}_{i} + d \hat{n}_{i} \hat{n}_{i+1} \right) + d \hat{N}+ \sum_{i\neq j} \omega^{i-j} \hat{n}_{i} \hat{n}_{j} \right] \Pi,\nonumber \\
\fl &=& \Pi  \frac{1}{\hat{N}^{2}}\left[ \sum_{i} \left(\hat{n}_{i} \hat{n}_{i} + d \hat {n}_{i} \hat{n}_{i+1} \right) + d \hat{N}+ \sum_{i=0}^{d-1} \left(\sum_{k=1}^{(d-1)/2} (\omega^{k}+\omega^{-k}) \hat{n}_{i} \hat{n}_{i+k}\right) \right] \Pi. \nonumber \\
\label{EQ:STOKES_d}
\end{eqnarray}
For $d=3$, as in this case $\omega+\omega^{-1}=-1$, the formula (\ref{EQ:STOKES_d}) reads
\begin{eqnarray}
\sum_{m=0}^3\abs{ \hat{R}_{m}}^2&=&\Pi \frac{1}{\hat{N}^{2}}  \left[ \sum_{i=0}^{2} \left(\hat{n}_{i} \hat{n}_{i} + 3 \hat{n}_{i} \hat{n}_{i+1} \right) + 3 \hat{N}+ \sum_{i=0}^{2} (\omega + \omega^{-1} ) \hat{n}_{i} \hat{n}_{i+1} \right] \Pi, \nonumber \\
&=& \Pi \frac{1}{\hat{N}^{2}} \left[ \sum_{i} \left(\hat{n}_{i} \hat{n}_{i} + 2 \hat{n}_{i} \hat{n}_{i+1} \right) + 3 \hat{N}\right] \Pi, \nonumber \\
&=& \Pi \frac{1}{\hat{N}^{2}} \left( \hat{N}^{2} + 3 \hat{N}\right) \Pi = \Pi + \Pi \frac{3}{\hat{N}} \Pi.
\end{eqnarray}
Therefore, we have  (\ref{EQ:STOKES_2}).

\section{Derivation of (\ref{EQ:LOWER_B})}
\label{APX:UBOUND_R}
Here we derive the relation (\ref{EQ:LOWER_B}) for any separable states, namely
\begin{eqnarray}
\label{EQ:LOWER_B_APX}
\sum_{m}\ave{\hat{R}_{m}^{A}\hat{R}^{\dagger B}_{m} + \hat{R}_{m}^{\dagger A}\hat{R}_{m}^{B}}_{\rm sep} \leq 2.
\end{eqnarray}
First we notice that as separable states are of the form of a convex combination $\varrho_{\rm sep}=\sum_{k}p_k\varrho_{k}^{A}\otimes \varrho_{k}^{B}$, we can search the maximum of LHS of (\ref{EQ:LOWER_B_APX}) using
\begin{eqnarray}
\max_{\varrho^A, \varrho^B} \sum_{m} \Tr \left[ \left(\hat{R}_{m}^{A}\hat{R}^{\dagger B}_{m} + \hat{R}_{m}^{\dagger A}\hat{R}_{m}^{B}\right) \varrho^{A} \otimes \varrho^{B} \right].
\end{eqnarray}
One can further simply the derivations by considering only tensor products of pure states, what we do further on. 
Note that
\begin{eqnarray}
&&\sum_{m} \Tr \left[ \left(\hat{R}_{m}^{A}\hat{R}^{\dagger B}_{m} + \hat{R}_{m}^{\dagger A}\hat{R}_{m}^{B}\right) \varrho^{A} \otimes \varrho^{B} \right] \nonumber \\
&=& \sum_{m} \left( \ave{\hat{R}_{m}^{A}}_{\varrho^A} \ave{\hat{R}_{m}^{\dagger B}}_{\varrho^B} +\ave{\hat{R}_{m}^{\dagger A}}_{\varrho^A} \ave{\hat{R}_{m}^{B}}_{\varrho^B} \right).
\end{eqnarray}
Thus in order to find the maximum we must know the general properties of $\ave{\hat{R}_{m}^{A}}_{\varrho^A}$ and $\ave{\hat{R}_{m}^{B}}_{\varrho^B}$.
Specifically, what will be needed will be the upper 
bound for 
\begin{eqnarray}
\sum_{m=0}^3  \langle\hat{R}_{m}^{X} \rangle_{\varrho^X}
\langle \hat{R}^{\dagger X}_{m}\rangle_{\varrho^X},
\end{eqnarray}
where $X=A,B$,
as by Cauchy inequality 
\begin{equation}
\left(\sum_{m} \Tr \left[ \left(\hat{R}_{m}^{A}\hat{R}^{\dagger B}_{m} + \hat{R}_{m}^{\dagger A}\hat{R}_{m}^{B}\right) \varrho^{A} \otimes \varrho^{B} \right]\right)^2 
\leq\max_X \left(4  \sum_{m=0}^3  |\langle{\hat{R}_{m}^{X}} \rangle_{\varrho^X}|^2\right).
\end{equation}

To this end we can use the algebraic relations already established in the derivation of (\ref{EQ:STOKES_2}), as given in (\ref{EQ:STOKES_d}). We take any pure state $|\psi\rangle$, and consider $\langle \hat{R}_m\rangle =\langle\psi| \hat{R}_m|\psi\rangle$. Notice that if one takes the average of 
$\sum_{m=0}^3 \hat{R}_{m}\hat{R}^\dagger_{m}$ with respect to $|\psi\rangle$ and inserts  $|\psi'\rangle \langle\psi'|=\Pi|\psi\rangle \langle\psi|\Pi$ between the  pairs of conjugated operators, one gets $\sum_{m=0}^3\langle \hat{R}_{m}\rangle \langle\hat{R}^\dagger_{m}\rangle$. This in turn be rearranged using the same algebraic steps of the first three of the equalities of (\ref{EQ:STOKES_d}), as they involve only the properties of $M(m)$ matrices. Below we put these manipulations explicitly for $d=3$: 
\begin{eqnarray}
\fl \sum_{m=0}^3\langle \hat{R}_{m}\rangle \langle\hat{R}^\dagger_{m}\rangle &=&
\sum_{i, j}\langle \psi'|  \frac{1}{\hat{N}} a^{\dagger}_i \left[ \left(\omega^{i-j} a_{i} |\psi'\rangle \langle\psi'| a_{j}^{\dagger} + \sum_{k} \omega^{k(i-j)} a_{i+1} |\psi'\rangle \langle\psi'| a_{j+1}^{\dagger}\right)  \right] a_j   \frac{1}{\hat{N}} |\psi'\rangle \nonumber \\
&=& \sum_{i, j} \omega^{i-j} \langle \psi'|\frac{1}{\hat{N}}  a^{\dagger}_i  a_{i} |\psi'\rangle \langle\psi'| a_{j}^{\dagger} a_j \frac{1}{\hat{N}}  |\psi'\rangle + 3 \sum_{i}\langle \psi'|\frac{1}{\hat{N}} a_i^\dagger 
a_{i+1}  |\psi'\rangle \langle\psi'|a_{i+1}^{\dagger} 
  a_i\frac{1}{\hat{N}} |\psi'\rangle \nonumber \\
&=&\sum_{i}\langle \psi'|\frac{1}{\hat{N}}  a^{\dagger}_i  \left(a_{i} |\psi'\rangle \langle\psi'| a_{i}^{\dagger}+ 3 a_{i+1}|\psi'\rangle \langle\psi'|
 a_{i+1}^{\dagger} \right) a_i  \frac{1}{\hat{N}}|\psi'\rangle \nonumber \\ 
 &+& \sum_{i\neq j} \omega^{i-j}\langle \psi'|\frac{1}{\hat{N}} a^\dagger_ia_{i} |\psi'\rangle \langle\psi'| a_{j}^{\dagger}  
 a_j \frac{1}{\hat{N}} |\psi'\rangle.
\end{eqnarray}
Thus, we get
\begin{eqnarray}
\fl \sum_{i}\langle \psi'|  \frac{\hat{n}_{i}}{\hat{N}}\ket{\psi'}\bra{\psi'} \frac{\hat{n}_{i}}{\hat{N}}\ket{\psi'} + 
3 \sum_{i}| \bra{\psi'} \frac{a_{i}^\dagger  a_{i+1}}{\hat{N}}\ket{\psi'}|^2 + \sum_{i\neq j} \omega^{i-j}  \bra{\psi'} \frac{\hat{n}_{i}}{\hat{N}} \ket{\psi'}\bra{\psi'} \frac{\hat{n}_{j}}{\hat{N}}|\psi'\rangle.
\end{eqnarray}
The upper bound of  middle  term is
\begin{eqnarray}
3 \sum_{i} \left \Vert \bra{\psi'} \frac{a_{i}^\dagger}{\sqrt{\hat{N}}} \right \Vert^2  \left \Vert \frac{a_{i+1}}{\sqrt{\hat{N}}}\ket{\psi'} \right \Vert^2
= 3 \sum_{i} \bra{\psi'} \frac{\hat{n}_{i}}{{\hat{N}}}\ket{\psi'}  \bra{\psi'} \frac{\hat{n}_{i+1}}{{\hat{N}}}\ket{\psi'},
\end{eqnarray}
while the last term is, due to the fact that for $d=3$ and for $i\neq j$ one has $\omega + \omega^{-1} = -1$, given by 
\begin{equation}
-\sum_{i}\bra{\psi'} \frac{\hat{n}_{i}}{{\hat{N}}}\ket{\psi'}  \bra{\psi'} \frac{\hat{n}_{i+1}}{{\hat{N}}}\ket{\psi'}.
\end{equation}
Thus we get 
\begin{equation}
\sum_{m=0}^3\langle \hat{R}_{m}\rangle \langle\hat{R}^\dagger_{m}\rangle \leq \big(\sum_{i}\langle \psi'|  \frac{\hat{n}_{i}}{\hat{N}}\ket{\psi'}\big)^2 = \langle \psi| \Pi\ket{\psi}^2 \leq 1.
\end{equation}


\section{Losses induced noise}
\label{APX:LOSS}
Here we study noise robustness of the entanglement criteria. As our model of noise we shall take losses of photon counts due to inefficiency of the detection. We shall assume that all detectors have the same efficiency.
 
In the case of a theoretical description of detection/collection losses,  an inefficient detector can be  emulated by a perfect one with a beam-splitter  of transmissivity $\eta$ in front of it, so that the probability that $m$ photons are counted while $n$ reach beam-splitter reads

\begin{equation}
p(m|n,\eta)=\left\{\begin{array}{ccc} \delta_{m,0}&&\eta=0\\ \delta_{m,n}&&\eta=1\\ \left(\begin{array}{c} n\\m\end{array}\right)\eta^m(1-\eta)^{n-m}&&\mathrm{otherwise}\end{array}\right.
\end{equation}

The numerical results to be presented here consider the six-mode bright squeezed vacuum state (\ref{EQ:BSV}):
\begin{eqnarray}\label{COMP}
\ket{\rm BSV}&=&\frac{1}{\cosh^3 \Gamma}\sum_{i=0}^{\infty}\sqrt{\frac{(i+1)(i+2)}{2}} \tanh^{i}\Gamma\ket{\psi^i},\nonumber\\
\ket{\psi^i}&=&\sqrt{\frac{2}{(i+1)(i+2)}}\sum_{i_1+i_2+i_3=i} \ket{i_1,i_2,i_3,i_1,i_2,i_3}.
\end{eqnarray}
We cut off the sequence at  to $i=10$. 
 With the losses model which we adopt one has 
\begin{eqnarray}\label{C3}
\fl \left\langle\left|\hat{R}_{m}^{A}-\hat{R}_{m}^{B}\right|^2\right\rangle_i
&=&\frac{2}{(i+1)(i+2)}\sum_{i_1+i_2+i_3=i}\sum_{a_{1}=0}^{i_1}\sum_{a_{2}=0}^{i_2}\sum_{a_{3}=0}^{i_3}\sum_{b_{1}=0}^{i_1}\sum_{b_{2}=0}^{i_2}\sum_{b_{3}=0}^{i_3}\nonumber\\
&\times&p(a_{1}|i_1,\eta)p(a_{2}|i_2,\eta)p(a_{3}|i_3,\eta) p(b_{1}|i_1,\eta)p(b_{2}|i_2,\eta)p(b_{3}|i_3,\eta)\nonumber\\
&\times&\left|\left(\frac{a_{1}+\omega a_{2}+\omega^2 a_{3}}{a_{1}+a_{2}+a_{3}}\right)'-\left(\frac{b_{1}+\omega b_{2}+\omega^2 b_{3}}{b_{1}+b_{2}+b_{3}}\right)'\right|^2,
\end{eqnarray}
where $\omega=\exp(2\pi \rmi/3)$ and $(\cdot)'$ implies that the fractions take value 0 when  the denominator is equal to zero. Here $\ave{\cdot}_i$ denotes the average over $\ket{\psi^i}$.  To calculate the right hand side of (\ref{EQ:ENT_COND}) we notice that 
\begin{eqnarray}
\label{C4}
\left\langle\Pi^X\frac{1}{\hat{N}^X}\Pi^X\right\rangle_i &=& \frac{2}{(i+1)(i+2)}\sum_{i_1+i_2+i_3=i} \sum_{x_{1}=0}^{i_1}\sum_{x_{2}=0}^{i_2}\sum_{x_{3}=0}^{i_3} \nonumber \\
&\times& p(x_{1} | i_1,\eta)p(x_{2} | i_2,\eta)p(x_{3} | i_3,\eta) \left(\frac{1}{x_{1}+x_{2}+x_{3}}\right)'.
\end{eqnarray}
Because of the symmetry of the state, and the detection efficiency, for each pair of conjugated unbiased bases $m$ one has the same value of  of $\left\langle\left|\hat{R}_{m}^{A}-\hat{R}_{m}^{B}\right|^2\right\rangle_i$.

These averages are summed with weights defined by the superposition coefficients:
\begin{eqnarray}
\fl \left\langle\left|\hat{R}_{m}^{A}-\hat{R}_{m}^{B}\right|^2\right\rangle_\Gamma&=&\frac{1}{\cosh^6 \Gamma}\sum_{i=0}^{10}\frac{(i+1)(i+2)}{2}\tanh^{2i}\Gamma\left\langle\left|\hat{R}_{m}^{A}-\hat{R}_{m}^{B}\right|^2\right\rangle_i,\nonumber\\
\fl \left\langle\Pi^X\frac{1}{\hat{N}^X}\Pi^X\right\rangle_\Gamma&=&\frac{1}{\cosh^6 \Gamma}\sum_{i=0}^{10}\frac{(i+1)(i+2)}{2}\tanh^{2i}\Gamma\left\langle\Pi^X\frac{1}{\hat{N}^X}\Pi^X\right\rangle_i,
\end{eqnarray}
where $\ave{\cdot}_{\Gamma}$ is the average over the state $\ket{\rm BSV}$.
Plugging this into inequality  (\ref{EQ:ENT_COND}) we get that if
\begin{equation}
\label{Crit13}
4\left\langle\left|\hat{R}_{m}^{A}-\hat{R}_{m}^{B}\right|^2\right\rangle_{\Gamma}-3\left\langle\Pi^A\frac{1}{\hat{N}^A}\Pi^A\right\rangle_{\Gamma}-3\left\langle\Pi^B\frac{1}{\hat{N}^B}\Pi^B\right\rangle_{\Gamma}<0,
\end{equation}
we detect entanglement. For large $\Gamma$  we find that the LHS of (\ref{Crit13}) is less than 0 for $\eta>0.154$. For $\Gamma \approx 0$, we get requirement of $\eta_{\rm crit}\approx 1/4$.

For the criterion based on intensities  (\ref{INTENSITIES}) for each $i$ one gets $\eta_{\rm crit}=1/4$. The formulas which enter the inequality are 
(\ref{C3}) with the denominators in primed expressions equal to 1 (removed), and similarly in (\ref{C4}) with the denominator and numerator interchanged.

\end{document}